\documentstyle[12pt]{article}
\begin{document}
\title{Decay of heavy-light hybrids in HQET sum rules}
\author {{\small Tao Huang$^{1,2}$\thanks{huangt@hptc5.ihep.ac.cn}, 
Hongying Jin$^2$\thanks{jhy@hptc5.ihep.ac.cn} and Ailin
Zhang$^2$\thanks{zhangal@hptc5.ihep.ac.cn}}\\
{\small $^1$ CCAST (World Laboratory), P. O. Box 8730, Beijing, 100080}\\ 
{\small $^2$ Institute of High Energy Physics, P. O. Box 918, Beijing, 100039, 
P. R.China}\\}
\date {}
\maketitle
\begin{center}
\begin{abstract}
The decay widths of the $0^{++}$ and $1^{-+}$ heavy-light hybrids to
B(D) and pion are calculated by using the QCD sum rules.
The interpolated current of the  hybrid  is chosen as
$g\bar q\gamma_{\alpha}G_{\alpha\mu}^aT^ah_{\it v}(x)$.
In order to simplify the calculation and avoid the ambiguity of three-point
correlation function, a two-point correlation function between the pion and
vacuum is used instead. The decay width of the $0^{++}\to B(D)$ is about
$12(16)$ MeV while the $1^{-+}\to B(D)$ is around $0.4(1.8)$ MeV. We keep
the leading order of $1/M_Q$ expansion in our calculation for convenience.
\end{abstract}
\end{center}
\section{Introduction}
\indent
\par It has been a long time for us to search the exotic hadrons such as
the glueballs and hybrids. No candidate that has been
found confirmedly so far. Recently, there are some evidences for the existence of
hybrids resonance however. The E852 collaboration at
BNL\cite{bnl} has reported a
$J^{pc}=1^{-+}$ isovector resonance $\hat{\rho}(1405)$  in the
reaction $\pi^-p\rightarrow\eta\pi^0n$  with the mass $1370\pm 16^{+50}_{-30}$
MeV and width  $385\pm 40^{+65}_{-105}$ MeV. Since the normal $\bar qq$
meson only has even spin in the channel ($\eta\pi^0$), the state
$1^{-+}$ must be beyond the quark model.
The Crystal Barrel collaboration has also claimed to find an evidence
 in $p\bar p$ annihilation which may be an resonance with a mass of
$1400\pm20\pm20$ MeV and a width of $310\pm50^{+50}_{-30}$ MeV\cite{bnl}.
Recently, the E852 collaboration put forth evidence for another
$J^{pc}=1^{-+}$ exotic meson $\hat{\rho}(1600)$\cite{e852}, decaying to $\rho\pi$, in
the reaction $\pi^-p\rightarrow\pi^+\pi^-\pi^-p$, with a mass and width of
$1593\pm8$ MeV and $168\pm20$ MeV respectively. 
If these experiments have been confirmed in the future, they will provide 
a very strong evidence for the existence of the constituent gluon. 

\par Theoretically, the hybrids have been studied widely by various 
methods such as bag
model\cite{bag}, flux-tube model\cite{ft}, QCD sum rules\cite{sum}, lattice
\cite{latt} and some other models\cite{other}. However, there are few works 
about the heavy-light hybrids except for QCD sum rules\cite{sum} in which the
spectrum of 
the heavy-light hybrid was given. Compared with  
 $\bar bb$ and $\bar cc$ hybrid, the heavy-light 
hybrid is easier to be dealt with since the heavy quark effective theory
(HQET) may work in such a system. As we know, HQET has led to much progress
in the theoretical understanding of the properties of hadrons\cite{hqet}. In
such a framework, many phenomenology methods become easier to be controlled.
For instance, QCD sum rule in HQET has been applied to various calculations
of physical parameters, such as the decay constants, form factors and decay
widths\cite{rules}.  
\par In the heavy-light hybrids case, the calculation in full QCD theory in
Ref. \cite{grw} shows that the constituent of gluon gives  
a contribution more than $1.0 GeV$ to the mass, so the "light freedom" 
is too heavy and the availability of $1/M_Q$ expansion seems 
problematic. Let's consider this problem. In the framework of
$1/M_Q$ expansion, 
the mass of heavy hadron can be expressed as 
\begin{equation}
M=M_Q+m_{light-freedom}+\lambda^2/M_Q+...,
\end{equation}

where the $\lambda$ have the dimension of mass and
can be estimated by the mass split of the doublet such as
($B$,$B^*$),($D$,$D^*$) because of the heavy quark spin symmetry.
For instance, $M_{B^*}-M_B=0.05$ GeV, then
$\lambda\approx 0.5 GeV$, which is roughly equal to
$m_{light-freedom}=M_B-M_b=0.5$ GeV if we choose $M_b=4.7 GeV$, therefore,
 the next-leading correction is around $\lambda/M_b\approx 1/10$.
Similarly, one can obtain
$\lambda\approx m_{light-freedom}=M_D-M_c=0.5 GeV$ if one chooses
$M_c=1.3 GeV$, and
$\lambda/M_c\approx 1/3$ for doublet ($D$,$D^*$). From the calculation in
Ref. \cite{grw}, the mass split
of $0^{++}$ and $1^{-+}$ is $0.5 GeV$ for $b$ hybrid,
so $\lambda=1.6 GeV\sim M_{b,1^{-+}}-M_b$ and the next-leading
correction is around $\lambda/m_b\approx 1/3$. For the $c$ hybrid,
the mass split from Ref. \cite{grw} is  $0.8 GeV$, $\lambda=1.0 GeV$, which
deviates a little more from $m_{light-freedom}= M_{c,1^{-+}}-m_c\approx 2.0 GeV$(it
is also a little smaller than $1.6$ GeV) , and
the next-leading correction $\lambda/M_c\sim 1.0$.
Comparing with B and D mesons , we can think it is still safe to use
$1/M_Q$ expansion in $b$ hybrid system while it should be very careful 
to apply HQET to $c$ hybrids.
\par As to the decay property of hybrids, the results given by the QCD
sum rules \cite{vg} gave a strong disfavor to the experiments. The decay
width for some individual channel calculated by them is much lower than the
reported experimental results. As we know, all the predictions in sum rules
before is dependent on the three-point correlation function, so the uncontrolled
ambiguity resulted from the double Borel transformation will be brought in
and the infrared divergence in the soft pion limit will appear too. To avoid
these disadvantages, we make use of the two-point correlation function 
between the
pion and vacuum in our derivation instead of the normal three-point
correlation function. Besides, it simplifies our calculation.   
\par In this paper, we employ the HQET sum rules to
calculate the decay width of the $0^{++}$ and $1^{-+}$ hybrids to $B(D)$ meson and
pion. For convenience, we keep only the leading order of $1/M_Q$
expansion in our calculation, which may make some deviation in the $c$
hybrid case. Our numerical results show that the decay width of $0^{++}\to
B(D)$ is about $12(16)$ MeV while the decay width of $1^{-+}\to B(D)$ is only
$0.4(1.8)$ MeV. 
\par The paper is organized as follows.  
The  analytic formalism of HQET sum rules for the decay of hybrid is given 
in Sec. 2. In Sec. 3, we calculate some pion's matrix element which is
necessary in the sum rules for the calculation of decay width. 
The numerical results of the decay width were obtained in Sec. 4. 
We give the conclusion and discussion in the last section. 

\section{HQET sum rules for the decay of heavy-light hybrid mesons}
\indent
\par In this paper, we consider the following processes
\begin{eqnarray}
\label{processa}
H_b(0^{++})(k) \longrightarrow B(0^{-+})(k-q) + {\it \pi}^{\pm}(q),\\
\label{processb}
H_b(1^{-+})(k) \longrightarrow B(0^{-+})(k-q) + {\it \pi}^{\pm}(q),
\end{eqnarray}
where the $H_b(0^{++})(k)$ represents the hybrid with $b$ quark and momentum
$k$, which has $J^{pc}=0^{++}$, $H_b(1^{-+})(k)$ represents the hybrid with
$b$ quark and momentum $k$, which has $J^{pc}=1^{-+}$. Here, the electric
charges of the mesons except for pion have not been written out
explicitly. The cases of $H_c(0^{++}) \rightarrow D {\it \pi}^{\pm}$ and 
$H_c(1^{-+}) \rightarrow D {\it \pi}^{\pm}$ are completely analogous. To calculate
the decay widths of these processes, we consider the following two-point 
correlator \begin{equation}\label{correlation}
A_{\nu}(\omega',\omega,v)=i\int dx e^{ikx}\langle {\it
\pi^{\pm}(q)}|T\{j_{1\nu}(x),j_2(0)\}|0
\rangle=A(\omega',\omega)v_{\nu}+B(\omega',\omega)(-q_{\nu}+q\cdot
vv_{\nu})
\end{equation}
 Where $j_{1\nu}(x)=g\bar q\gamma_{\mu}G^a_{\mu\nu}T^ah_{\it v}(x)$,
$j_2(x)=\bar h_{\it v}\gamma_5 q(x)$, $h_{\it v}$ is the heavy quark
effective field with four velocity $v$, $A(\omega',\omega)$ and
$B(\omega',\omega)$
are scalar functions of $\omega$ and $\omega'$, where $\omega$ and $\omega'$
are defined as
\begin{eqnarray}\label{var}
\omega=2k\cdot v &,&\omega'=2(k-q)\cdot v.
\end{eqnarray}

$A(\omega',\omega)$ and $B(\omega',\omega)$ 
 are determined through the spectral density saturated by the mesons
corresponding to the interpolated currents, respectively. Here
we mention that the interpolated current for
the hybrid with a fixed $J^{pc}$ is not unique, different interpolated 
current corresponds to different state\cite{sum}.    
\par Before carrying out the operator product expansion, we can simplize
the correlation function Eq. (\ref{correlation}) firstly. Since the free heavy quark
propagator in $x$ representation in HQET is ${\it \int_{0}^{\infty} d\tau\delta(x-v\tau)\frac {1+\rlap/v}{2}}$
and the interaction of the heavy quark with the gluon field $A_\mu$ in the
leading order of $1/M_Q$ expansion is $g\bar hv\cdot Ah$. Then in the 
fixed-point gauge $x_{\mu} A_{\mu}=0$(which will be used throughout this
paper), the full propagator of the heavy
quark $\langle 0|T(h(x)\bar h(0)|0 \rangle$ in the leading order of
$1/M_Q$ expansion is the same as the free one.
 Then the correlator (\ref{correlation}) reduces to the one including only light
quarks. 
\par Performing the OPE in  $A_{\nu}$, we get
\begin{eqnarray}\label{cor}
A_{\nu}&=&i\int dx e^{ikx}\langle {\it \pi^{\pm}(q)}|T\{j_{1\nu}(x),j_2(0)\}|0\rangle
\\\nonumber
&=&-\int dx e^{i(k-q)x}\int d\tau \delta (x-v\tau) Tr\{\Gamma \langle {\it
\pi^{\pm}(q)} |G_{\mu\nu}(0)q(-x)\bar q(0) |0\rangle\}\\\nonumber
&=&-\int dx e^{i\omega'\tau/2} Tr\{\Gamma\{\langle {\it \pi^{\pm}(q)}
|G_{\mu\nu}q\bar q |0\rangle
-\langle {\it \pi^{\pm}(q)}| \tau v\cdot D q\bar qG_{\mu\nu} |0\rangle\}\} ,
\end{eqnarray}
where $\Gamma=\gamma_\mu \frac{1+\rlap/v}{2}\gamma_5$ and 
$G_{\mu\nu}=igG^a_{\mu\nu}T^a$. In deriving the
last equation in (\ref{cor}), the quark field $q(-x)$ is expanded around
zero and only the first two terms are kept, the numerical results in 
section. 4 assure it a good expansion.  We stress here that, 
since the perturbative theory does not break chiral symmetry
$\partial_\mu j^{\pi}_\mu = 0$, there is no perturbative
contribution to the sum rules in (\ref{cor}) in the chiral limit.  
The two matrix elements in (\ref{cor}) will be obtained through another
suitable sum rules with the method used in Ref. \cite{element}. 
Their detailed expressions will be given in the next section.
\par On the other side, $A(\omega',\omega)$ and $B(\omega',\omega)$ can be
represented with the
spectral density through the dispersion relations.
 Because the perturbative contribution vanish, we
suppose that the contribution of continuum states vanishes too. 

From the definition of Eq. (\ref{var}), we have
\begin{eqnarray*}
\omega-\omega'=2q\cdot v
\end{eqnarray*}
and one has to make the double Borel transformation on the two variables
$\omega$ and $\omega'$. The soft pion approximation was used in the spectral
density in Ref. \cite{rules}, which deduced functions of single variable.
However, in the
case of heavy-light hybrid's decay, the soft pion approximation is not good
since the pion momentum may be large. Here we make a more reasonable
approximation
\begin{eqnarray}\label{app}
\omega-\omega'=2q\cdot v \approx 2(\Lambda-\Lambda')
\end{eqnarray}
in the infinite heavy quark mass limit, where $\Lambda \sim m_H-M_Q$ and
$\Lambda' \sim m_{meson}-M_Q$. Eq. (\ref{app}) can be obtained from
\begin{eqnarray*}
k^2+q^2-(k-q)^2=2k\cdot q \approx 2M_Q q\cdot v
\end{eqnarray*}
as $M_Q\to\infty$. We find that the difference between $\omega$ and
$\omega'$ is around $1-2$ GeV in the case of heavy-light hybrid. With the
single pole terms and double pole term included in the spectral density, then 
$A(\omega',\omega)$ and $B(\omega',\omega)$ are expressed as functions of
the single variable $\omega'$, respectively 
\begin{eqnarray}
\label{a}
A(\omega')=\frac{f_{H^+}f_mg_{H^+m\pi}m^4_{H^+}m^2_m}{(2\Lambda'-\omega')^2M^3_Q}+{c_0
\over {2\Lambda'-\omega'}},\\
B(\omega')=\frac
{f_{H^-}f_mg_{H^-m\pi}m^4_{H^-}m^2_m}{(2\Lambda'-\omega')^2M^3_Q}+{c_1 \over
{2\Lambda'-\omega'}},
\label{b}
\end{eqnarray}
where $m_H$, $m_m$ and $M_Q$ are the mass of the hybrid, meson,
and heavy quark, respectively. $c_0$ and $c_1$ in the above equations are
constants. $f_i$ are decay constants and $g_{H^\pm m\pi}$ refers to decay
amplitudes. For convenience, they are defined in full QCD theory as below 
\begin{eqnarray}
\langle 0|j_H|H(0^{++})\rangle =f_{H^+}m^3_{H^+}k_\nu & ,
& \langle 0|j_H|H(1^{-+})\rangle =f_{H^-}m^4_{H^-}\epsilon_\nu ,\\\nonumber
\langle 0|j_D| D\rangle = -if_Dm^2_D/M_c & , 
& \langle 0|j_B| B\rangle = -if_Bm^2_B/M_b ,\\\nonumber
\langle \pi^{\pm} (q) D|H(0^{++})\rangle =g_{H^+D\pi} & , 
&\langle \pi^{\pm} (q) D|H(1^{-+})\rangle =g_{H^-D\pi}\epsilon\cdot q ,\\\nonumber
\langle \pi^{\pm} (q) B|H(0^{++})\rangle =g_{H^+B\pi} & , 
&\langle \pi^{\pm} (q) B|H(1^{-+})\rangle =g_{H^-B\pi}\epsilon\cdot q .
\end{eqnarray}

Therefore, only one variable Borel transformation is needed.
 Taking the Borel transformation 
\begin{equation}
\frac {1}{\tau}\hat B^{\omega'}_\tau = \!\!\lim_{\matrix{ n\to\infty \cr
 -\omega'\to\infty \cr \tau=-\omega'/\tau fixed}}\!\!
 \frac {\omega'^n}{\Gamma (n)}(-\frac
{d}{d\omega'})^n 
\end{equation}
on both sides of the sum rules and differentiate the results in the variable
$1/\tau$, we get the expression for the decay
amplitudes
\begin{eqnarray}\label{am1}
g_{H^+m\pi} ={M^3_Q\over
f_{H^+}f_mm^4_{H^+}m^2_m}[2\Lambda' A'(\tau)+A_0]e^{2\Lambda'/\tau},\\
g_{H^-m\pi} ={M^3_Q\over
f_{H^-}f_mm^4_{H^-}m^2_m}[2\Lambda' B'(\tau)+B_0]e^{2\Lambda'/\tau}\label{am2},
\end{eqnarray}
where the $A'(\tau)$ and $B'(\tau)$ are the Borel transformed function of
$A(\omega')$ and $B(\omega')$, respectively. Their expressions are obtained  
\begin{eqnarray}\label{ab}
A'(\tau)&=&4\sqrt{2}\{[3b_1-2(m_H-m_m)^2d_2]+[(m_H-m_m)F_1/4+F_3/16]{1\over\tau}\},\\
B'(\tau)&=&4\sqrt{2}[2(m_H-m_m)d_2+F_1/(12\tau)],
\end{eqnarray}
and
\begin{eqnarray*}
A_0&=&4\sqrt{2}[(m_H-m_m)F_1/4+F_3/16],\\
B_0&=&{\sqrt{2}F_1 \over 3} ,
\end{eqnarray*}
where the parameters $b_i$,$d_i$ and $F_i$ will be given in the next section. 
\par Once the decay amplitudes are determined, it's straightforward to
obtain the decay widths of the processes (\ref{processa}) and (\ref{processb}).  

\section{The sum rules for pion's matrix elements} 
\indent
\par The key to the sum rules is the determination of the two pion's matrix elements in
(\ref{cor}). The  first pion's matrix element
$\langle {\it \pi^{\pm}(q)} |G_{\mu\nu}q \bar q |0\rangle $ has already been
given in Ref. \cite{rules}(d) as
\begin{eqnarray}
\langle\pi^i(q)|D_\mu D_\nu q^a_\alpha(0)\bar q^b_\beta(0)|0\rangle&=&
i\{[g_{\mu\nu}(a_1+
a_2\rlap/q)-i\sigma_{\mu\nu}b_1+i\varepsilon_{\mu\nu\rho\sigma}
\gamma^{\rho}q^{\sigma}
\gamma_5b_2\nonumber\\&&\mbox{}+(q_\mu\gamma_\nu+q_\nu\gamma_\mu)c_1
+(q_\mu\gamma_\nu-q_\nu\gamma_\mu)d_1\nonumber\\&&\mbox{}
+i(q_\mu\sigma_{\lambda\nu}+q_\nu\sigma_{\lambda\mu})q^\lambda c_2
+i(q_\mu\sigma_{\lambda\nu}-q_\nu\sigma_{\lambda\mu})q^\lambda d_2\nonumber
\\&&\mbox{}
+(e_1+e_2\rlap/q)q_\mu q_\nu]\gamma_5\}_{\alpha\beta}
\left({\tau_i\over 2}\right)_{ab}\;
\end{eqnarray} 
where the definition and the numerical results of the coefficients can be
found in the same Ref..
From the Lorentz covariance, the matrix element
$\langle \pi(q)^{\pm}|(D_\alpha q)^i_j\bar qG_{\beta\gamma}|0\rangle$ can be
written as
\begin{eqnarray}
\langle \pi^{\pm}(q)|D_\alpha q\bar qG_{\beta\gamma}|0\rangle=& 
\{if_1q_\alpha \sigma_{\beta\gamma} - f_2\gamma_\alpha [q_\beta \gamma_\gamma-
q_\gamma\gamma_\beta]+if_3\gamma_\alpha \sigma_{\beta\gamma} -
f_4(g_{\alpha\beta}q_\gamma-g_{\alpha\gamma}q_\beta)\\\nonumber
&-f_5(g_{\alpha\beta}\gamma_\gamma-g_{\alpha\gamma}\gamma_\beta)\}\gamma_5
+if_6\epsilon_{\alpha\beta\gamma\rho}q_\rho-f_7q_\alpha
(q_\beta\gamma_\gamma- q_\gamma \gamma_\beta) ,
\end{eqnarray}
where $f_i$ are some constants to be determined. By using the motion equation 
in the chiral limit and performing some special traces, we can obtain the
following equations about $f_i$
\begin{eqnarray}
f_1=-{1\over 4}F_1,&f_2={1\over 12}(F_1-F_2),&f_3={1\over 48}F_3,\\\nonumber
f_4={1\over 12}F_1-{1\over 3}F_2,&f_5=-{1\over 24}F_3,&f_6={1\over 4}F_1,
\end{eqnarray}
and $f_7$ vanishes. Where $F_1$, $F_2$ and $F_3$ are some constants defined as
\begin{eqnarray}
\langle \pi^{\pm}(q)|\bar qG_{\mu\nu}D_\alpha q|0\rangle &=&
g\varepsilon_{\mu\nu\alpha\rho}q_\rho F_1 ,\\
\langle \pi^{\pm}(q)|\bar q\gamma_5G_{\mu\nu}D_\alpha q|0\rangle &=& 
ig(g_{\alpha\nu}q_\mu - g_{\alpha\mu}q_\nu)F_2,\\
\langle \pi^{\pm}(q)|\bar q\gamma_5\gamma_\mu G_{\mu\nu}D_\nu q|0\rangle &=& 
igF_3.
\end{eqnarray}

In order to get the numerical results of the constants $F_i$, we take
advantage of some other suitable sum rules, through which the constants
$F_i$ can be obtained by the following correlation functions, respectively
\begin{eqnarray}
\Pi^1_{\mu\nu\alpha}(q) &=& i\int dx e^{iq\cdot x}\langle 0|T\{\bar qG^a_{\mu\nu}T^aD_\alpha q(x),
\bar q\gamma_5 q(0)\}|0\rangle ,\\\nonumber
\Pi^2_{\mu\nu\alpha}(q) &=& i\int dx e^{iq\cdot x}\langle 0|T\{\bar
q\gamma_5G^a_{\mu\nu}T^aD_\alpha q(x),
\bar q\gamma_5 q(0)\}|0\rangle ,\\\nonumber
\Pi^3_\rho (q) &=& i\int dx e^{iq\cdot x}\langle 0|T\{\bar
q\gamma_5\gamma_\mu G^a_{\mu\nu}T^aD_\nu q(x),
\bar q\gamma_\rho\gamma_5q(0)\}|0\rangle.
\end{eqnarray}
where the $\bar q$ or $q$ refer to the u, d quark or antiquark corresponding
to which in the $\pi^{\pm}$.
\par Keeping the matrix elements of OPE up to dimension six and omiting the radiative correction,
we obtain the Borel transformed functions
\begin{eqnarray}
F_1(M)&=&\{-{2\alpha_s \over 9(4\pi)^3}(M^2)^2 + {1\over 24}\langle
0|{\alpha_s \over \pi}G^2 |0\rangle + {4\pi \over 27M^2}\alpha_s \langle
0|\bar qq|0\rangle ^2\}{(m_u+m_d)M^2 \over f_\pi m^2_\pi}e^{m^2_\pi/M^2},\\\nonumber
F_2(M)&=&-F_1(M),\\\nonumber
F_3(M)&=&\{{-31\alpha_s \over 15(4\pi)^3}(M^2)^2 + {1 \over 4}\langle
0|{\alpha_s \over \pi}G^2 |0\rangle - {4\pi \over 9M^2}\alpha_s \langle
0|\bar qq|0\rangle ^2\}{M^2 \over f_\pi}e^{m^2_\pi/M^2}. 
\end{eqnarray}
where $M$ is the Borel transform parameter of the sum rules.
\par To obtain the numerical results, the constants and condensates are chosen as
\begin{eqnarray}
f_{\pi} = 132 MeV ,& m_{\pi} = 140 MeV ,& m_u+m_d = 15 MeV ,\\\nonumber
\langle 0|{\alpha_s \over \pi}G^2|0 \rangle = 0.012 GeV^4 ,& \langle 0|\bar
qq|0 \rangle = -(0.24 GeV)^3 ,&\alpha_s \langle 0|\bar qq|0 \rangle ^2 =
8\cdot 10^{-5} GeV^6 
\end{eqnarray}
and where the $\alpha_s=0.4$. 
\par The numerical results are showed as Fig. 1 and Fig. 2, where one can
find stable platforms in the region $M\sim 1.1-1.6$ GeV. they read   
\begin{eqnarray}
F_1=-F_2=4.0\cdot 10^{-3} GeV^4 &,&F_3=2.3\cdot 10^{-2} GeV^5.
\end{eqnarray}

\section{the numerical results of the decay width} 

\indent
\par Before going on the numerical calculation of the decay
width, there are still some parameters necessary to be fixed.
The heavy quark and meson masses are given as those in Ref. \cite{epj} and the decay
constants of B and D mesons are chosen as Ref. \cite{rr}. The masses of the
heavy-light hybrids have been computed in Ref.\cite{grw} in the full
QCD theory though the b quark mass($4.23$ GeV) and some other parameters
were chosen unsuitablely, we just recite the results of
their calculation in this paper.
\par The decay constants of hybrids hadn't been
determined in Ref. \cite{grw} yet, so we fix them through the
parameters and formulae given in it as
\begin{eqnarray}
f^2_i & = &{M^2 \over m^8_R} e^{m^2_R \over M^2}\Pi_i (M^2),
\end{eqnarray}
where $\Pi_i (M^2)$ refer to the functions corresponding to the $J^{pc}=0^{++}$
and $J^{pc}=1^{-+}$ states
\begin{eqnarray}
\Pi_s (M^2) & = &{3\alpha_s \over 4\pi^3}M^6\int_{0}^{1}dx \int_{0}^{1}dy
x(1-x)y^3(3y-2)e^{-{m^2 \over M^2xy}}\\\nonumber
& - &{1 \over 8}\langle {\alpha_s \over \pi}G^2 \rangle M^2\int_{0}^{1}dx
x^2e^{-{m^2 \over M^2x}} - {\alpha_s \over \pi} \langle m{\bar qq} \rangle M^2 \int_{0}^{1}dx
x(1-x)e^{-{m^2 \over M^2x}},\\\nonumber
\Pi_v (M^2) & = &{3\alpha_s \over 4\pi^3}M^6\int_{0}^{1}dx \int_{0}^{1}dy
x(1-x)y^4e^{-{m^2 \over M^2xy}}\\\nonumber
& + &{1 \over 24}\langle {\alpha_s \over \pi}G^2 \rangle M^2\int_{0}^{1}dx
x(2-x)e^{-{m^2 \over M^2x}} - {\alpha_s \over 3\pi} \langle m{\bar qq} \rangle M^2 \int_{0}^{1}dx
x(1-x)e^{-{m^2 \over M^2x}}.
\end{eqnarray}
where $m_R$ is the mass of the hybrids, the Borel parameter $M$ is chosen
in the region where both the masses and the decay constants have platform
and the scale of the $\alpha_s$ is set at the Borel variable. Their results
are given in the table 1 too.
\par With the help of Eq. (\ref{am1}) and Eq. (\ref{am2}), the numerical results of the decay amplitudes
are shown as Fig. 3 and Fig. 4. The platform is not good, we choose the
value around the region $\tau \sim 2.5$ GeV, which is the suitable region in
sum rules analysis, as our results. The numerical result shows that the next-leading
contribution is less than $1/5$ of the leading term at this region, so the OPE
is well performed. All the parameters and some results calculated are collected in table. 1.
\begin{table}
\caption{\label{tab1}
{\it Parameters input and the decay amplitudes.}}
\begin{center}
\begin{tabular}{|c||c|c|c|c|c|}
\hline
&&&&&\\
\hspace{2.0cm}& $M_Q$ & $m_m$ & $f_m$ & $m_H(0^{++})$ & $m_H(1^{-+})$ \\
\hline
&&&&&\\
b quark & 4.7 GeV & 5.28 GeV & 180 MeV & 6.8 GeV & 6.3 GeV \\
\hline
&&&&&\\
c quark & 1.3 GeV & 1.87 GeV & 190 MeV & 4.0 GeV & 3.2 GeV \\
\hline
\hline
&&&&&\\
\hspace{2.0cm} & $f_{H(0^{++})}$ & $f_{H(1^{-+})}$ & 
$g_{H^+m\pi}$& $g_{H^-m\pi}$ &~~~~~~\\
\hline
&&&&&\\
b quark & 0.0012 & 0.0012 & 3.2 GeV & 1.2 &~~~~~~\\
\hline
&&&&&\\
c quark & 0.004 & 0.004 & 2.0 GeV & 1.1 &~~~~~~\\
&&&&&\\
\hline
\end{tabular}
\\
\end{center}
\end{table}
 
\par  As to the processes (\ref{processa}) and (\ref{processb}) we
considered, with the help of the formulae for two body decay, the decay
widths in the leading order of $1/M_Q$ expansion are given as
\begin{eqnarray}\label{de}
\Gamma(H(0^{++})\rightarrow m(0^{-+}) + {\it \pi}) =&
{g^2_{Hm\pi} \over 8\pi}{|q| \over m^2_H} =&
{g^2_{Hm\pi}\over 16\pi}{m^2_H-m^2_m \over m^3_H},\\\nonumber
\Gamma(H(1^{-+})\rightarrow m(0^{-+}) + {\it \pi}) =&
{g^2_{Hm\pi} \over 24\pi}{|q|^3 \over m^2_H} =&
{(m^2_H-m^2_m)^3g_{Hm\pi}^2\over 192 \pi m^5_H}.
\end{eqnarray}

When all the numerical results are put in, the decay width corresponding
to processes (\ref{processa}) and (\ref{processb}) are obtained 
\begin{eqnarray}
\Gamma(H(0^{++})\rightarrow B(0^{-+}) + {\it \pi}) = 12 MeV ,
&\Gamma(H(0^{++})\rightarrow D(0^{-+}) + {\it \pi}) = 16 MeV ,\\
\Gamma(H(1^{-+})\rightarrow B(0^{-+}) + {\it \pi}) = 0.4 MeV ,
&\Gamma(H(1^{-+})\rightarrow D(0^{-+}) + {\it \pi}) = 1.8 MeV .
\end{eqnarray}

As we know, the decay of hybrids appears to follow the almost universal
selection rule that gluonic excitations cannot transfer angular momentum to
the final states as relative angular momentum, that is to say, the decay of
hybrids to $S + P$ final pairs of mesons are preferred, while the decay of
hybrids to two S-wave mesons are suppressed \cite{page}. But the selection
rule is not absolute, in the flux tube and constituent glue models, it can
be broken by wave function and relativistic effects, and the bag model predict
that it is also possible that the excited quark loses its angular momentum
to orbital angular momentum \cite{godfrey}. Though the results obtained here
are much larger than those calculated for the similar decay modes in
the light hybrids case \cite{vg}, which depended on the construction of
three point correlation function, these results couldn't be excluded by
theory.    
\par The reason that the decay widths of the $1^{-+}$ hybrids are much smaller
than those of the $0^{++}$ hybrids results from   
the additional factor $q_\pi^2/m_H^2$ in the width formulae(\ref{de}) 
compared with that of the $0^{++}$'s, since in the decay channel 
$H_{b,c}(1^{-+})\rightarrow B(D)\pi$, the final states are in a relative
P-wave.

\section{Conclusion and Discussion}
\indent
\par In this paper, we discuss the decay of
the heavy-light hybrids $H(0^{++})$ and $H(1^{-+})$ to
$B(D){\it \pi}$ by  QCD sum rule approach in the framework of
HEQT. With the help of the spectrum of the heavy-light hybrids\cite{grw}, we
argue, in the $b$ hybrid case, convergence of $1/M_Q$ expansion
is similar as that in the $D$ meson case, while it is not so good for the $c$
hybrid. The two-point correlation function instead of the normal three-point
correlation function is employed for the derivation, and the heavy quark freedom
is extracted out of the correlation functions in the $x$ representation,
which leads to the estimate of some pion's matrix elements and
facilitates the calculation. In the leading order of $1/M_Q$ expansion, the process of the calculation becomes much
easier.  
\par To find the hybrids, one approach is to look for an excess of observed
state over the number predicted by the quark model, the other is to search
for quantum numbers which cannot be accommodated in the quark model, so the
experiments place main efforts on the exotic mesons' searching and the
current experiments concern mainly the light hybrids sector, however,
it's necessary to extend the search to the heavy-light hybrids sector.
The calculation shows that the decay width of
$H(0^{++})\rightarrow B(D){\it \pi^+}$ is
around $12(16)$ MeV, while the decay width
$H(1^{-+})\rightarrow B(D){\it \pi^+}$ is only $0.4(1.8)$ MeV.
If people think the total width of hybrid is around 200 MeV,
the channel $B(D)\pi$ may not be
dominant in hybrid's decay, for instance, the channels
$B,D(B^*,D^*)\eta'$ may be
larger because of QCD anomaly. The calculation of these and other channels
is beyond this paper. 
\par There are some uncertainties unavoidable in this paper, which result
from such as pole approximation, $1/M_Q$ expansion(we keep only the leading 
order term) and OPE(we keep only the first two terms of OPE in
(\ref{cor})), omitting the higher orders would give some 
uncertainties. Besides, the calculation is dependent on the numerical
results of the masses and decay constants of the hybrids taken
from the Ref. \cite{grw}, which were determined by
some unpopular parameters. To follow the calculation of Ref.
\cite{grw} with the new values of the parameters is beyond the goal of this
paper. Certainly, it's necessary to identify the spectrum and decay width of
heavy-light hybrids through other models and other sum rules, for example,
light cone sum rules. Only when most of the models give detailed research on
hybrids, can we declaim that we have a good understanding about the hybrids.

\vspace{1.0cm}
{\bf Acknowledgment}

This work is supported in part by the national natural science foundation 
of P. R. China.

\newpage
\par
{\huge\bf Figure caption}\\
\par
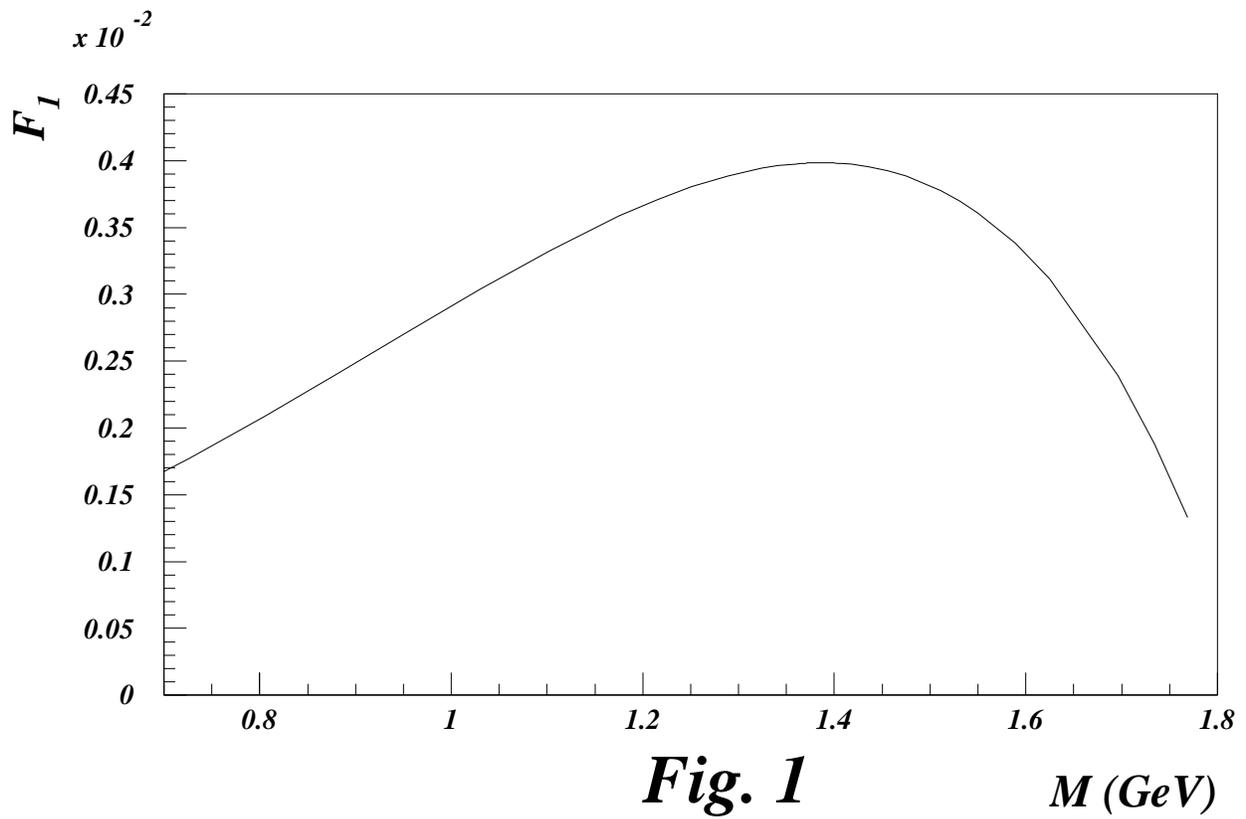
Figure 1: $F_1$ versus Borel variable M.\\
\par
\par
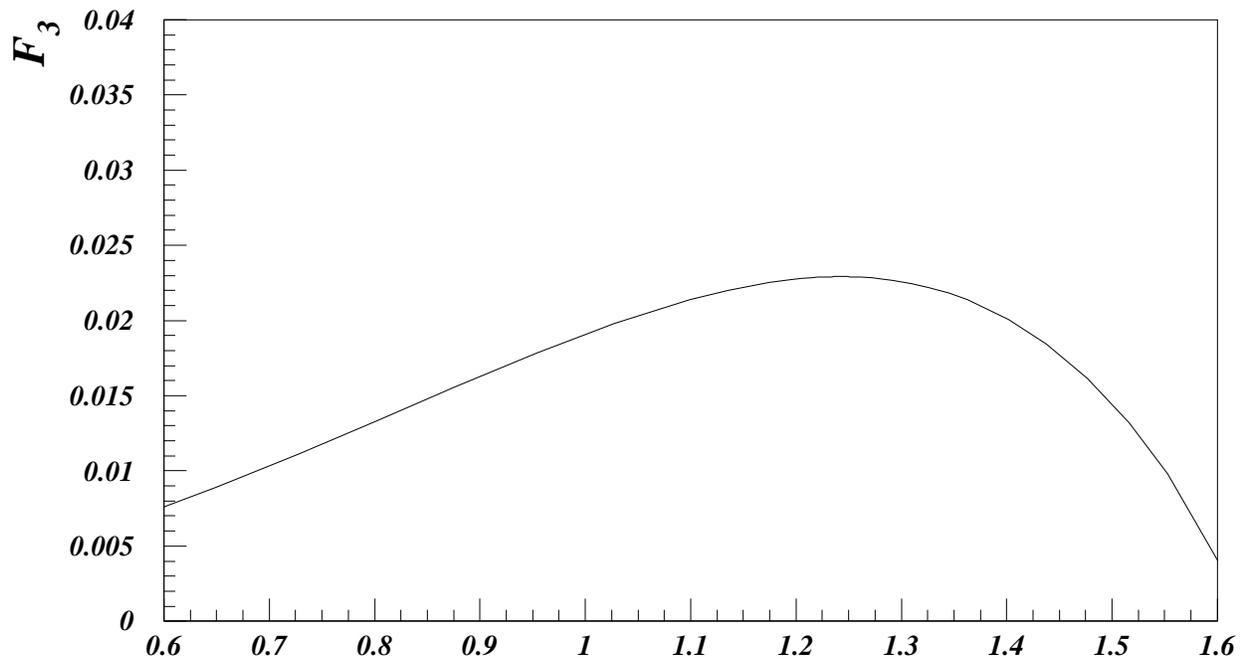
Figure 2: $F_3$ versus Borel variable M.\\
\par
\par
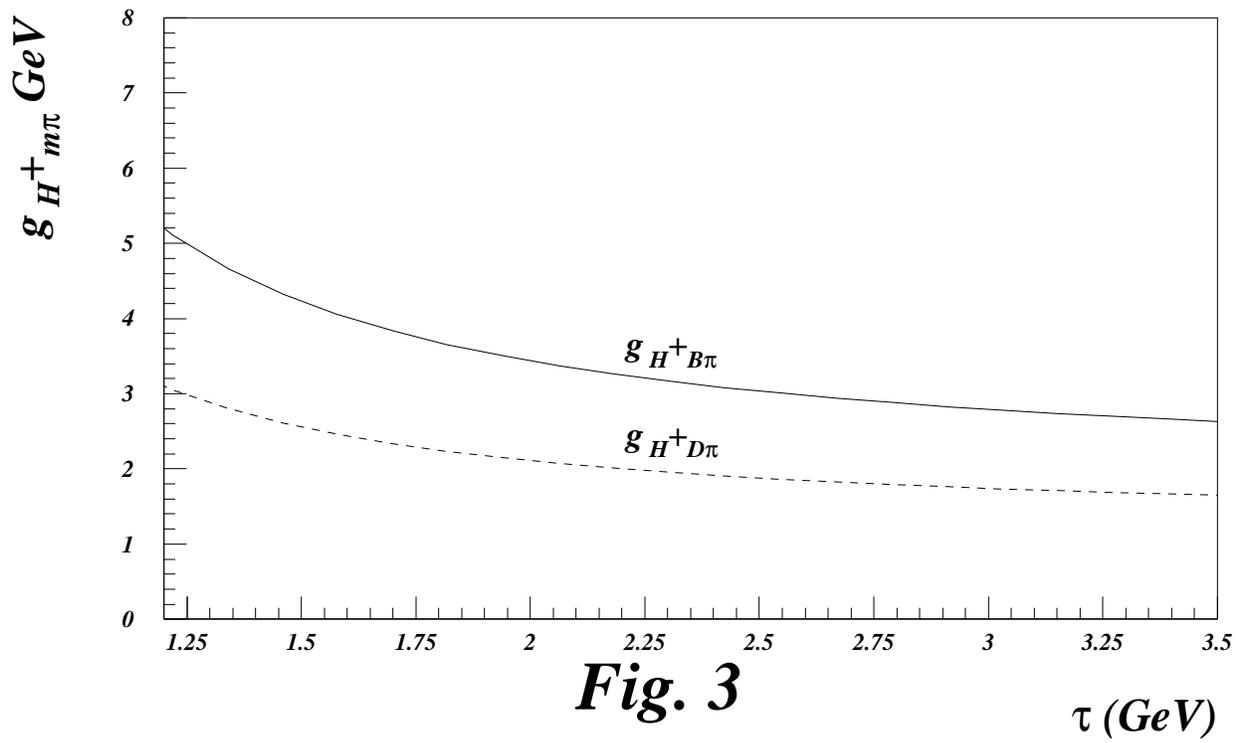
Figure 3: Decay amplitude of $0^{++}$ heavy-light hybrids versus Borel
variable $\tau$.\\
\par
\par
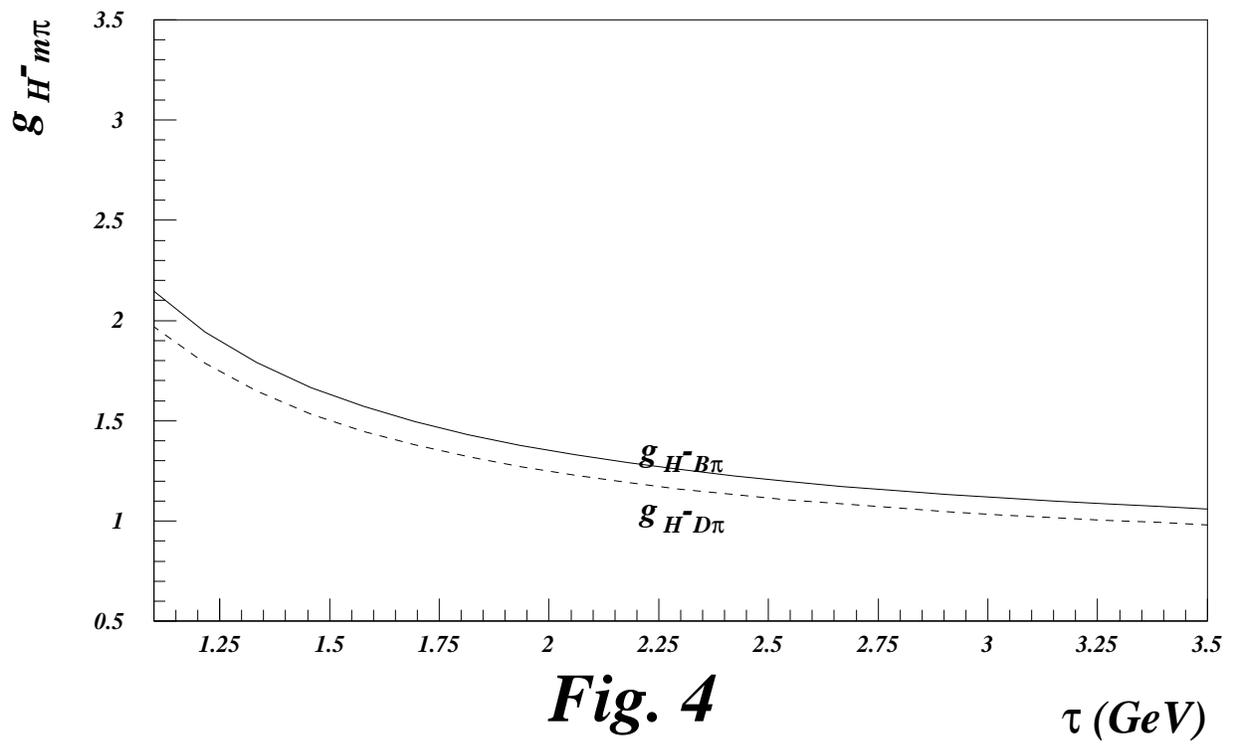
Figure 4: Decay amplitude of $1^{-+}$ heavy-light hybrids versus Borel
variable $\tau$.\\

\end{document}